\documentclass[9pt,twocolumn,twoside]{osajnl}
\usepackage{textcomp}

\journal{ol} 
\setboolean{shortarticle}{true}

\title{300~GHz generation based on a Kerr microresonator frequency comb stabilized to a low noise microwave reference}

\author[1,*]{Tomohiro Tetsumoto}
\author[2]{Fumiya Ayano}
\author[1]{Mark Yeo}
\author[2]{Julian Webber}
\author[2]{Tadao Nagatsuma}
\author[1]{Antoine Rolland}

\affil[1]{IMRA America Inc., Boulder Research Labs, 1551 South Sunset St, Suite C, Longmont, Colorado 80501, USA}
\affil[2]{Graduate School of Engineering Science, Osaka University, 1-3 Machikaneyama, Toyonaka, Osaka 560-8531, Japan}

\affil[*]{Corresponding author: ttetsumo@imra.com}

\begin{abstract}
In this letter, we experimentally demonstrate low noise 300~GHz wave generation based on a Kerr microresonator frequency comb operating in soliton regime. The spectral purity of a 10~GHz GPS-disciplined dielectric resonant oscillator is transferred to the 300~GHz repetition rate frequency of the soliton comb through an optoelectronic phase-locked loop. Two adjacent comb lines beat on a uni-travelling carrier photodiode emitting the 300~GHz millimeter-wave signal into a waveguide. In an out-of-loop measurement we have measured the 300~GHz power spectral density of phase noise to be -88~dBc/Hz, -105~dBc/Hz at 10~kHz, 1~MHz Fourier frequency, respectively. The free-running fractional frequency instability at 300~GHz is $1 \times 10^{-9}$ at 1~second averaging time. Stabilized to a GPS signal, we report an in-loop residual instability of $2 \times 10^{-15}$ at 1~second which averages down to < $1 \times 10^{-17}$ at 1000~seconds. Such system provides a promising path to the realization of compact, low power consumption millimeter-wave oscillators with low noise performance for out-of-the-lab applications.
\end{abstract}
\setboolean{displaycopyright}{false}

\begin{document}
\maketitle
Millimeter-wave and TeraHertz generation with state-of-the-art noise performance in chip-scale form factor will undeniably stimulate a plethora of civilian and defense applications including high-resolution radar~\cite{Li:14}, non-destructive imaging~\cite{Elsa:2019}, 5G and 6G cellular network deployment~\cite{Calva:19}, wireless communication~\cite{nagatsuma2016advances,yi2019300}, global navigation system~\cite{Tang:13}, satellite communication~\cite{Vizard:06} as well as fundamental science with applications such as radioastronomy~\cite{Shillue:13,Nand:11}, and rotational spectroscopy~\cite{Wang:18,Wang:18_2}. While millimeter-wave oscillators based on complementary metal-oxide-semiconductor (CMOS) technology have shown spectacular performance in terms of size and power consumption, power spectral density of phase noise performance is lacking~\cite{nikpaik2017219}.
Photonics sources circumvent electronic noise and bandwidth limitations, however, size and usability outside of the lab are still fairly arguable. Soliton formation in Kerr microresonator frequency comb (microcomb) is a promising approach in order to use advantages of the photonic sources in a chip-scale form factor. Noise performance of a soliton microcomb is mainly due to the frequency noise of the pump continuous-wave (CW) pump laser and the thermo-reflective noise of the resonator~\cite{yi2017single,Huang:19,Liao:17}. Optical frequency division of an ultra-stable CW laser to the millimeter-wave domain would be the most performant approach to achieve ultra-low noise millimeter-wave signals but the complexity of the system could not be ignored as the soliton microcomb needs to be self-referenced which as of today is not trivial to realize in chip-scale form factor. Recently, a microresonator based optical frequency comb was used to generate a low noise and stable millimeter-wave signal at 300~GHz, however the experimental setup requires the use of a bulky microwave synthesizer and an hydrogen maser~\cite{zhang2019terahertz}. Alternatively, in this letter, we propose to transfer the spectral purity of a low noise GPS-disciplined 10~GHz dielectric resonant oscillator (DRO) to the 300~GHz repetition rate frequency of the soliton microcomb. An electro-optic (EO) frequency comb generator, based on three cascaded optical phase modulators, is used to bridge the 10~GHz DRO and the 300~GHz repetition rate frequency, here detected with solely two adjacent comb lines. The resulting power spectral density of phase noise at 300~GHz is $-88$~dBc/Hz, $-105$~dBc/Hz at 10~kHz, 1~MHz Fourier frequency, respectively, ruled by the phase noise multiplication of the DRO from 10~GHz to 300~GHz. As the 300~GHz spectral purity is directly derived from the DRO, disciplining it to a GPS signal also stabilizes the long-term drift of the 300~GHz signal. While the free-running fractional frequency stability is $1 \times 10^{-9}$, stabilization to the GPS signal reduces the residual in-loop fractional instability to $2 \times 10^{-15}$. 

\begin{figure}[htbp]
\centering
\fbox{\includegraphics[width=\linewidth]{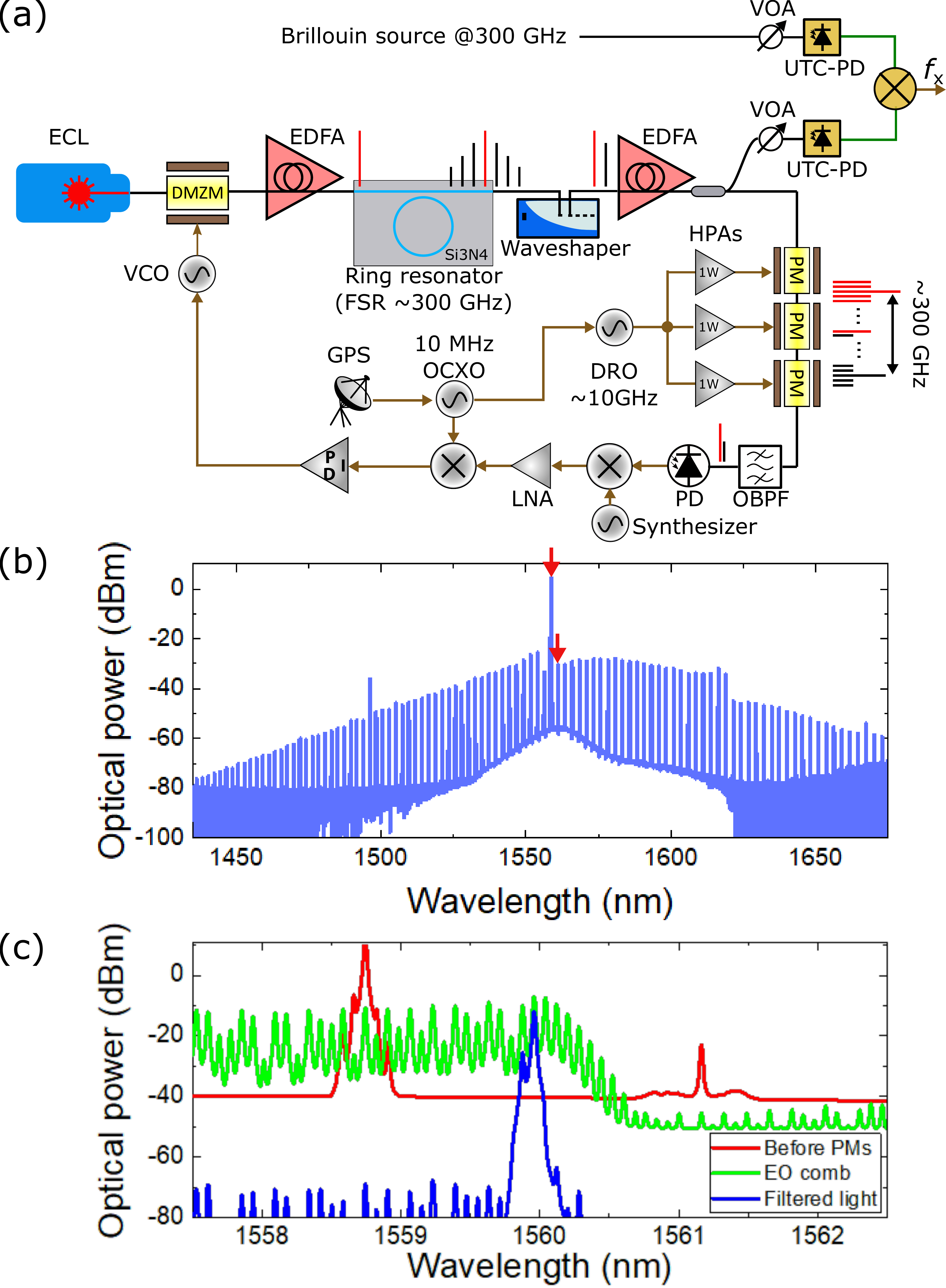}}
\caption{(a) Experimental setup for microcomb-based 300~GHz generation. See text for details. VOA: variable optical attenuator, HPAs: high power amplifiers. (b) Optical spectrum of the generated microcomb. Two lines used for the millimeter-wave generation are indicated by red arrows. (c) Optical spectrum of the EO comb of the selected two lines. The spectrum shape of the EO comb is controlled so that the beat signal of the filtered light is maximized by fine-tuning the DRO frequency (phase effectively).}
\label{fig:1}
\end{figure}

Figure~\ref{fig:1}(a) shows a schematic of the experimental setup for the low noise millimeter-wave generation. A 1560~nm continuous wave optical light from an external cavity laser (ECL) is amplified with an erbium doped fiber amplifier (EDFA) and coupled as a pump with a silicon nitride ring resonator with free spectrum range (FSR) of about 300~GHz. A microcomb is initiated by a fast pump frequency sweep through single side band modulation with a dual-parallel Mach–Zehnder modulator (DMZM) \cite{briles2018interlocking, kuse2019control}. Optical spectrum of the generated microcomb is shown in Fig.~\ref{fig:1}(b), where the sech$^2$ shaped envelop of the comb lines with a single FSR frequency spacing suggests that the microcomb forms a single soliton. A microresonator-based optical frequency comb operating in the soliton regime guarantees phase coherence between optical lines and formation of optical pulses. However, as the goal of this work is to generate a millimeter-wave signal we choose to opt out of the use of the optical pulse train and to simply select two adjacent optical comb lines so we do not saturate the photodetector. Therefore, two neighbouring optical lines of the microcomb are spectrally filtered with a waveshaper, amplified with an EDFA and split into two paths (the selected two lines are indicated by red arrows in Fig.~\ref{fig:1}(b)). Light on one arm is sent to an uni-traveling-carrier photodiode (UTC-PD), where millimeter-wave radiation with frequency equals to the microcomb's FSR of about 300~GHz is generated through the photomixing of the two optical lines. Through a variable optical attenuator 10~mW impinges on the UTC photodiode. With an average photocurrent of 4~mA a millimeter-wave power of 40~\textmu W at 300 GHz is generated. 

\noindent
While it is possible to increase the optical power to get to a photocurrent of 10~mA emitting about 0.2~mW at 300~GHz, in the early stage of this experiment it is preferable to only use half of the UTC photodiode capability for long-term and robust operation. 

While the two optical lines are inherently phase coherent when the microcomb is operated in soliton regime, the pump laser frequency noise and the thermo-refractive noise are ruling the spectral purity of the 300~GHz signal. The carrier envelope offset frequency $f_\mathrm{ceo}$ does not directly affect the millimeter-wave signal as it cancels out at the photodetection when two optical modes with mode indices of $n+1$ and $n$ beat on the UTC detector:
\begin{equation}
    ((n+1)\times f_\mathrm{rep} + f_\mathrm{ceo})-(n \times f_\mathrm{rep} + f_\mathrm{ceo}) = f_\mathrm{rep}.
\end{equation}
\noindent
The repetition rate fluctuations $\delta f_\mathrm{rep}$ are then directly affecting the fluctuations of the millimeter-wave signal $\delta f_\mathrm{mmw}$. In our present case, one only needs to phase lock the repetition rate as the reference lies in the microwave domain. The frequency comb is then basically used as a frequency multiplier. As the frequency of the 10~GHz reference and the 300~GHz millimeter-wave are about 5 octaves away, one needs a mechanism to down-convert the millimeter-wave signal to the microwave domain. This can be realized with a sub-harmonic mixer. However, residual phase noise and conversion loss can be very large and the a faithful down-conversion of the millimeter-wave signal to the microwave domain is not guaranteed, in particular when it comes to realize this function in a chip-scale form factor. Down-conversion using optoelectronic components have shown great potential, however, bridging important gaps between millimeter-wave and THz signals and the microwave signals can be challenging in a chip-scale form factor. Here, we chose to realize the down-conversion with optoelectronic components. The down-conversion scheme and the feedback loop implementation are depicted on Fig.~\ref{fig:1}(a). The two-wavelengths light, at the output of the waveshaper and EDFA (see red optical spectrum on Fig.~\ref{fig:1}(c)), splits to a lower arm connected to three cascaded phase modulators (PM) driven by the 10~GHz DRO. The DRO is synchronized to a 10~MHz signal of an oven-controlled crystal oscillator (OCXO) phase-locked to a GPS signal (very slow feedback bandwidth of about 100~Hz is applied to the voltage bias of the DRO in order to keep its very good spectral purity while being disciplined to the GPS signal). Phase modulation of the two-wavelength optical signal generates side bands from the two optical lines (see green optical spectrum on Fig.~\ref{fig:1}(c)). As the EO combs are driven at 10~GHz and that both wavelengths are separated by 300~GHz, both EO combs need to span at least 300~GHz and harmonics $\pm 15^{\mathrm{th}}$, respectively, have to spectrally overlap within a few GHz. As can be observed on the blue spectrum on Fig.~\ref{fig:1}(c), the $\pm 15^{\mathrm{th}}$ side bands are picked out with a tunable optical band path filter (OBPF) with a minimum 3-dB bandwidth as low as 6.25~GHz (Fig.~\ref{fig:1}(c)). We maximize the optical power selected by the OBPF by optimizing the DRO frequency and its output microwave power. This step is fundamental in obtaining a strong enough signal-to-noise ratio. The light is detected with a photodiode (PD) and an intermediate frequency $f_\mathrm{IF}$ around 1.2~GHz is photodetected, whose frequency is down-converted to 10~MHz level through a difference frequency mixing with a signal from a low noise synthesizer. The phase noise of this RF wave $\delta f_\mathrm{RF}$ is described as follows,
\setlength\abovedisplayskip{3pt}
\begin{equation}
\label{eq1}
    \delta f_\mathrm{RF} = \delta f_\mathrm{rep} - 2k\delta f_\mathrm{DRO},
\setlength\belowdisplayskip{3pt}
\end{equation}
\noindent
where $\delta f_\mathrm{rep}$ is the repetition rate phase noise, $\delta f_\mathrm{DRO}$ is DRO phase noise and $k$ is index for the order of the side bands \cite{rolland2011non,del2012hybrid}. The phase noise of the synthesizer signal used for the down-conversion is ignored since it is sufficiently small. After RF amplification of the signal with a low noise amplifier (LNA), an error signal is generated with a phase detector where the RF frequency is mixed with the same 10~MHz signal that is used as a reference of the DRO. The error signal is applied to the DMZM through a proportional-integral-derivative (PID) filter driving a voltage controlled oscillator (VCO) in order to nullify the phase difference between $f_\mathrm{rep}$ and $2kf_\mathrm{DRO}$. Once the feedback loop is closed, the phase noise of the repetition rate becomes a copy of the phase noise from the DRO of $2k\delta f_\mathrm{DRO}=30\delta f_\mathrm{DRO}$ following Eq.~\ref{eq1}, within the phase-locked loop feedback bandwidth (here at most 1~MHz mainly depending on the signal-to-noise ratio of the intermediate frequency). The frequency multiplication factor of 30 will then induce a phase noise increase of $20 \times \log(30) = 29.5$~dB. Note that this noise relationship only holds for frequency components lower than the feedback bandwidth.

\begin{figure}[htbp]
\centering
\fbox{\includegraphics[width=\linewidth]{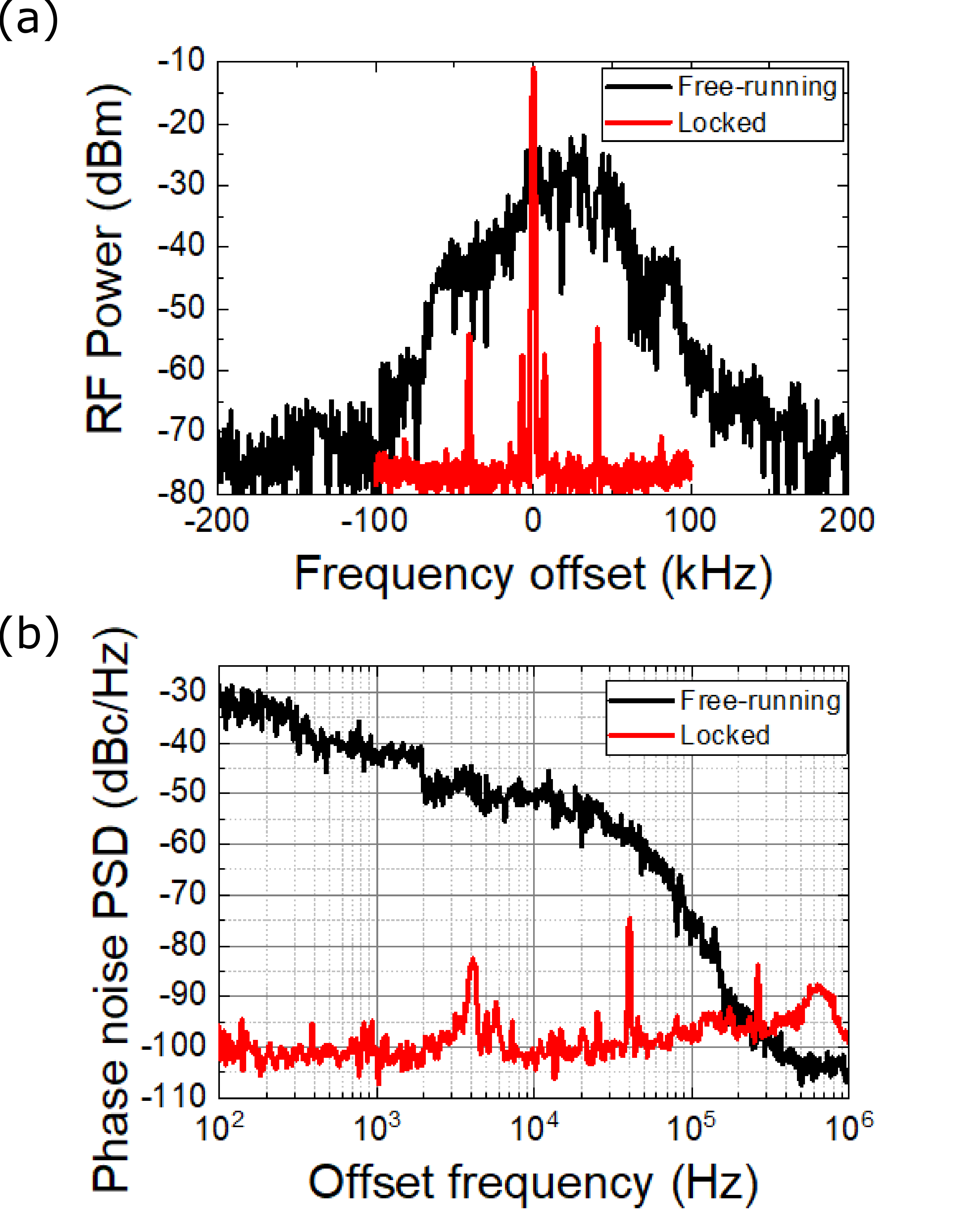}}
\caption{(a) In-loop RF spectrum. Horizontal axis shows frequency offset from a peak. Resolution bandwidth is 1~kHz. (b) Phase noise of in-loop signals.}
\label{fig:2}
\end{figure}

Figure~\ref{fig:2}(a) shows RF spectrum of the in-loop signal. The in-loop signal is picked up between the low noise amplifier and the RF frequency mixer. The detected RF spectrum is noisy in the free-running condition and a coherent peak cannot be resolved. Indeed, from the RF spectrum, the spectrum becomes sharp and stable with the PID control. This transition is also observed clearly in plots of power spectrum density (PSD) of phase noise as shown in Fig.~\ref{fig:2}(b). The free-running noise is dramatically suppressed below offset frequencies lower than the feedback bandwidth.

\begin{figure}[htbp]
\centering
\fbox{\includegraphics[width=\linewidth]{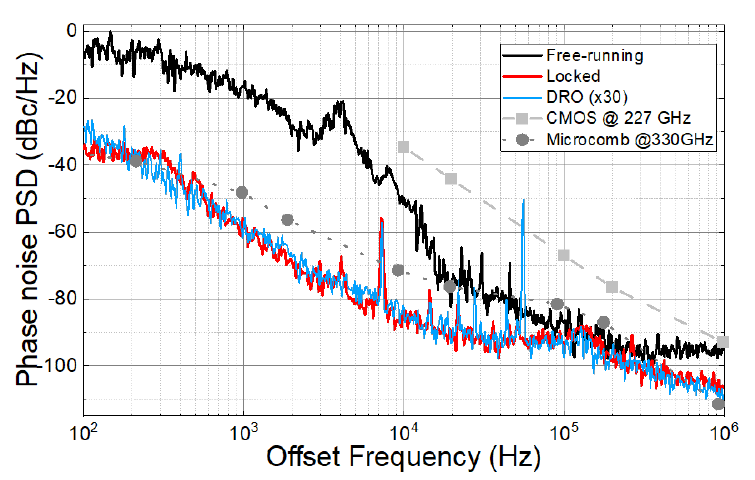}}
\caption{Measured phase noise of out-of-loop signals. Solid lines are obtained in this work, and dashed and dotted lines show approximate noise level of millimeter-wave sources in relating works \cite{nikpaik2017219, zhang2019terahertz}.}
\label{fig:3}
\end{figure}

In order to verify that the spectral purity of the DRO is indeed faithfully transferred to the soliton comb repetition rate, an out-of-loop measurement needs to be performed. While, it would be possible to compare the soliton comb based millimeter-wave signal to a multiplied microwave synthesizer using a sub-harmonic mixer, we decided to perform the measurement using a second millimeter-wave source based on photonic technologies. The reference source is built around a fiber ring cavity generating two Stokes waves induced by stimulated Brillouin scattering. The details of this source and its performance is described in the following letter \cite{li2019low}. The two millimeter-wave-sources are then mixed on a millimeter-wave fundamental mixer generating a electrical signal around 1~GHz. The millimeter-wave reference based on Brillouin scattering is sent to the LO port of the frequency mixer. For optimized operation of the mixer, we input 200~\textmu W of millimeter-wave power (corresponding to an average photocurrent of 9~mA) to the LO port. The 300~GHz signal derived from the microcomb is input to the RF port and yields a similar 9~mA average photocurrent. Phase noise of out-of-loop signal is shown in Fig~\ref{fig:3}. The free-running phase noise is reduced to the noise floor determined by the DRO noise in the locked conditions which indicates perfect spectral purity transfer from 10~GHz to 300~GHz. The obtained phase noise of $-88$~dBc/Hz ($-106$~dBc/Hz) at 10~kHz (1~MHz) is about 53~dB (12~dB) smaller than that of a low noise CMOS oscillator based on electrical multiplexing \cite{nikpaik2017219} (dashed line in Fig.~\ref{fig:3}). On the other hand, the phase noise is about 16~dB (13~dB) better at 10~kHz (100~kHz) compared to the microcomb-based 300~GHz source in \cite{zhang2019terahertz} whereas the phase noise of $-106$~dBc/Hz at 1~MHz is about 5~dB worse than it.

\begin{figure}[htbp]
\centering
\fbox{\includegraphics[width=\linewidth]{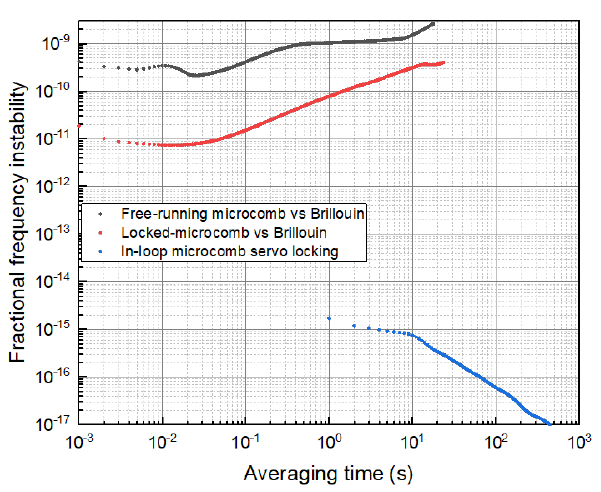}}
\caption{Measured fractional frequency instability of in- and out-of-loop signals.}
\label{fig:4}
\end{figure}

Finally, we characterize frequency stability of the source by using a frequency counter and the result is depicted in Fig.~\ref{fig:4}. We obtain fractional frequency instability of $8\times 10^{-16}$ at 1~second gate time for the in-loop signal while that for free-running is $1\times 10^{-9}$. This indicates that the frequency instability of comb's repetition rate is significantly suppressed by the PID control. On the other hand, we obtain a fractional frequency instability of $2\times 10^{-11}$ for the out-of-loop signal, which is only one order of magnitude better than that of the free-running. This is due to the instability of the reference 300~GHz source used for the down-conversion where no active mechanism for frequency stabilization is operating. We believe the actual out-of-loop instability will be much lower than that measured since there is no critical source of the instability in the system. Indeed, in-loop fractional frequency measurement suggests that the setup in this state would follow a GPS signal faithfully. Here, we steer a Rubidium microwave clock with a 1 pulse per second signal generated with a GPS  antenna. This could result in approximately, a fractional instability of $1 \times 10^{-11}$ and < $1\times 10^{-12}$ at 1~second and 10000~seconds averaging time, respectively.

In summary, we demonstrated low noise millimeter-wave generation at 300~GHz through photomixing of two lines of a microcomb stabilized to a DRO synchronized to a GPS signal. The measured phase noise of $-88$~dBc/Hz at 10~kHz is the lowest for oscillators based on chip-scale technologies. We demonstrate that the phase noise reported is limited by the phase of the DRO itself. The in-loop fractional frequency instability of $2\times 10^{-15}$ at 1~s indicates large reduction of the instability of the comb repetition rate and suggests a high stability of the source while it is not confirmed in the out-of-loop signal measurement due to the instability of the reference millimeter-wave source. The experiment is performed with equipment which can be realized in a compact manner. Also, the method can be applied for generation of other millimeter-wave frequency with a resonator with different FSR or by employing two non-adjacent comb lines yielding millimeter-wave with a frequency that is a harmonics of the FSR. This may require broader EO combs and higher photodetector bandwidth to bridge the larger frequency difference.
In order to demonstrate a prospective chip-scale millimeter-wave oscillator, one could argue that electro-optic comb generation is not promising. However, the down-conversion mechanism can be replaced with other components such as an injection-locked diode laser comb \cite{bustos2019direct} operating near 10~GHz.
This work shows a great potential of a microcomb as a low noise and compact millimeter-wave oscillator, whose features are attractive, in particular, for mobile wireless communication as well as radio-astronomy where the reported technology, in this letter, could meet the requirement of transferring the stability of microwave atomic clocks in the millimeter-wave domain. 

\medskip
\noindent\textbf{Acknowledgement} The authors thank Yuzuru Uehara for technical supports and Scott Papp from NIST for providing the silicon nitride microresonator chips. They are grateful to Martin Fermann for fruitful discussions at the early stage of this work.\\

\noindent\textbf{Disclosures} The authors declare no conflicts of interest.

\bibliography{bib_2020microcomb_locked_to_DRO.bib}

\begin{thebibliography}{10}
\newcommand{\enquote}[1]{``#1''}

\bibitem{Li:14}
Y.~Li, A.~Rashidinejad, J.-M. Wun, D.~E. Leaird, J.-W. Shi, and A.~M. Weiner,
  {\protect\JournalTitle{Optica}} \textbf{1}, 446 (2014).

\bibitem{Elsa:2019}
M.~{ELsaadouny}, J.~{Barowski}, J.~{Jebramcik}, and I.~{Rolfes},
  \enquote{Millimeter {Wave SAR Imaging for the Non-Destructive Testing of
  3D-printed Samples},} in \emph{2019 International Conference on
  Electromagnetics in Advanced Applications (ICEAA),}  (2019), pp. 1283--1285.

\bibitem{Calva:19}
E.~{Calvanese Strinati}, S.~{Barbarossa}, J.~L. {Gonzalez-Jimenez},
  D.~{Ktenas}, N.~{Cassiau}, L.~{Maret}, and C.~{Dehos},
  {\protect\JournalTitle{IEEE Vehicular Technology Magazine}} \textbf{14}, 42
  (2019).

\bibitem{nagatsuma2016advances}
T.~Nagatsuma, G.~Ducournau, and C.~C. Renaud, {\protect\JournalTitle{Nature
  Photonics}} \textbf{10}, 371 (2016).

\bibitem{yi2019300}
L.~Yi, K.~Iwamoto, T.~Yamamoto, F.~Ayano, Y.~Li, A.~Rolland, N.~Kuse,
  M.~Fermann, and T.~Nagatsuma, \enquote{300-{GHz}-band wireless communication
  using a low phase noise photonic source,} in \emph{2019 49th European
  Microwave Conference (EuMC),}  (IEEE, 2019), pp. 816--819.

\bibitem{Tang:13}
A.~{Tang} and Q.~{Gu}, \enquote{A high-precision millimeter-wave navigation
  system for indoor and urban environment autonomous vehicles,} in \emph{2013
  IEEE MTT-S International Microwave Symposium Digest (MTT),}  (2013), pp.
  1--3.

\bibitem{Vizard:06}
D.~Vizard, {\protect\JournalTitle{Microwave Journal}} \textbf{49}, 22 (2006).

\bibitem{Shillue:13}
W.~{Shillue}, W.~{Grammer}, C.~{Jacques}, H.~{Meadows}, J.~{Castro},
  J.~{Banda}, R.~{Treacy}, Y.~{Masui}, R.~{Brito}, P.~{Huggard}, B.~{Ellison},
  J.~{Cliché}, S.~{Ayotte}, A.~{Babin}, F.~{Costin}, C.~{Latrasse},
  F.~{Pelletier}, M.~{Picard}, M.~{Poulin}, and P.~{Poulin}, \enquote{A
  high-precision tunable millimeter-wave photonic {LO} reference for the {ALMA}
  telescope,} in \emph{2013 IEEE MTT-S International Microwave Symposium Digest
  (MTT),}  (2013), pp. 1--4.

\bibitem{Nand:11}
N.~R. {Nand}, J.~G. {Hartnett}, E.~N. {Ivanov}, and G.~{Santarelli},
  {\protect\JournalTitle{IEEE Transactions on Microwave Theory and Techniques}}
  \textbf{59}, 2978 (2011).

\bibitem{Wang:18}
C.~Wang, X.~Yi, J.~Mawdsley, M.~Kim, Z.~Wang, and R.~Han,
  {\protect\JournalTitle{Nature Electronics}} \textbf{1}, 421 (2018).

\bibitem{Wang:18_2}
C.~Wang, X.~Yi, J.~Mawdsley, M.~Kim, Z.~Hu, Y.~Zhang, B.~Perkins, and R.~Han,
  {\protect\JournalTitle{IEEE Journal of Solid-State Circuits}} \textbf{54},
  914 (2018).

\bibitem{nikpaik2017219}
A.~Nikpaik, A.~H.~M. Shirazi, A.~Nabavi, S.~Mirabbasi, and S.~Shekhar,
  {\protect\JournalTitle{IEEE Journal of Solid-State Circuits}} \textbf{53},
  389 (2017).

\bibitem{yi2017single}
X.~Yi, Q.-F. Yang, X.~Zhang, K.~Y. Yang, X.~Li, and K.~Vahala,
  {\protect\JournalTitle{Nature communications}} \textbf{8}, 1 (2017).

\bibitem{Huang:19}
G.~Huang, E.~Lucas, J.~Liu, A.~S. Raja, G.~Lihachev, M.~L. Gorodetsky, N.~J.
  Engelsen, and T.~J. Kippenberg, {\protect\JournalTitle{Physical Review A}}
  \textbf{99}, 061801 (2019).

\bibitem{Liao:17}
P.~Liao, C.~Bao, A.~Kordts, M.~Karpov, M.~H.~P. Pfeiffer, L.~Zhang,
  A.~Mohajerin-Ariaei, Y.~Cao, A.~Almaiman, M.~Ziyadi, S.~R. Wilkinson, M.~Tur,
  T.~J. Kippenberg, and A.~E. Willner, {\protect\JournalTitle{Optics Letters}}
  \textbf{42}, 779 (2017).

\bibitem{zhang2019terahertz}
S.~Zhang, J.~M. Silver, X.~Shang, L.~Del~Bino, N.~M. Ridler, and P.~Del’Haye,
  {\protect\JournalTitle{Optics Express}} \textbf{27}, 35257 (2019).

\bibitem{briles2018interlocking}
T.~C. Briles, J.~R. Stone, T.~E. Drake, D.~T. Spencer, C.~Fredrick, Q.~Li,
  D.~Westly, B.~Ilic, K.~Srinivasan, S.~A. Diddams \emph{et~al.},
  {\protect\JournalTitle{Optics Letters}} \textbf{43}, 2933 (2018).

\bibitem{kuse2019control}
N.~Kuse, T.~C. Briles, S.~B. Papp, and M.~E. Fermann,
  {\protect\JournalTitle{Optics Express}} \textbf{27}, 3873 (2019).

\bibitem{rolland2011non}
A.~Rolland, G.~Loas, M.~Brunel, L.~Frein, M.~Vallet, and M.~Alouini,
  {\protect\JournalTitle{Optics Express}} \textbf{19}, 17944 (2011).

\bibitem{del2012hybrid}
P.~Del’Haye, S.~B. Papp, and S.~A. Diddams, {\protect\JournalTitle{Physical
  Review Letters}} \textbf{109}, 263901 (2012).

\bibitem{li2019low}
Y.~Li, A.~Rolland, K.~Iwamoto, N.~Kuse, M.~Fermann, and T.~Nagatsuma,
  {\protect\JournalTitle{Optics Letters}} \textbf{44}, 359 (2019).

\bibitem{bustos2019direct}
R.~Bustos-Ramirez, L.~Trask, A.~Bhardwaj, G.~E. Hoefler, F.~A. Kish, and P.~J.
  Delfyett, \enquote{{Direct Optical Link between a mmWave Optical Frequency
  Comb and a Chip-Scale Mode-Locked Laser},} in \emph{2019 IEEE Photonics
  Conference (IPC),}  (IEEE), pp. 1--2.

\end{thebibliography}


\end{document}